\documentclass{Interspeech2024}

\interspeechcameraready 

\usepackage{xcolor}

\newcommand{\leo}[1]{{#1}}

\newcommand{\meanerror}[2]{#1\tiny$\pm$#2}

\title{EnCodecMAE: leveraging neural codecs for \\universal audio representation learning}

\name[affiliation={1,2}]{Leonardo}{Pepino}
\name[affiliation={1,2}]{Pablo}{Riera}
\name[affiliation={1}]{Luciana}{Ferrer}

\address{
  $^1$Instituto de Investigaci\'on en Ciencias de la Computaci\'on (ICC), CONICET-UBA, Argentina\\
  $^2$Departamento de Computaci\'on, FCEyN, Universidad de Buenos Aires (UBA), Argentina}
\email{lpepino@dc.uba.ar}

\begin{document}

\maketitle

\begin{abstract}
The goal of universal audio representation learning is to obtain foundational models that can be used for a variety of downstream tasks involving speech, music and environmental sounds. 
To approach this problem, methods inspired by works on self-supervised learning for NLP, like BERT, or computer vision, like masked autoencoders (MAE), are often adapted to the audio domain. 

In this work, we propose masking representations of the audio signal, and training a MAE to reconstruct the masked segments. The reconstruction is done by predicting the discrete units generated by EnCodec, a neural audio codec, from the unmasked inputs.
We evaluate this approach, which we call EnCodecMAE, on a wide range of tasks involving speech, music and environmental sounds. Our best model outperforms various state-of-the-art audio representation models in terms of global performance. Additionally, we evaluate the resulting representations in the challenging task of automatic speech recognition (ASR), obtaining decent results and paving the way for a universal audio representation.
\end{abstract}

\section{Introduction}
\label{sec:intro}

The search for universal audio models that can be used to solve a wide range of tasks related not only to speech, but also to music and environmental sounds, has received increased attention in recent years \cite{gong2022ssast, wang2022towards, seth2023slicer, huang2022amae, baade2022mae, chong2023masked, chen2023beats}. Some works pretrain these models in a supervised way using existing large labeled datasets like Audioset \cite{koutini2022passt, gong21b_interspeech}. Other works use self-supervised learning (SSL), which is an effective way to leverage large unlabeled audio datasets. SSL methods resort to learning a pretext task, like masked language modelling, which consists of predicting masked parts of an utterance \cite{hsu2021hubert, discretebert} or contrasting the output on a masked time step with the correct output and a set of distractors taken from the same utterance \cite{baevski2020wav2vec}. These pretext tasks operate at the frame level and learn relationships between parts of an utterance. Contrastive learning has also been used to learn instance-level embeddings, by creating positive examples for a certain audio through augmentation techniques or by taking temporally-close fragments, and negative examples by extracting fragments from different audio samples \cite{jansen2018unsupervised, shor2020towards}. BYOL-A \cite{niizumi2021byol} is another recently proposed method to learn instance-level representations of audio signals without requiring negative examples, by minimizing the distance between two augmented views of a spectrogram encoded by a teacher and a student network.

The pretrained upstream models can be adapted to downstream tasks using relatively small labelled datasets by feeding the activations from one or more of its layers to a downstream network.
The pretrained model can be frozen \cite{wang2022towards, turian2022hear} or fine-tuned \cite{gong2022ssast, huang2022amae, chen2023beats}. 
When the task of interest has instance-level labels but the pretrained model generates frame-level representations, an operation that pools over time is applied to generate instance-level representations \cite{gong2022ssast, wang2022towards, huang2022amae}. 

Our work follows the ideas originally proposed in BERT~\cite{devlin2018bert} for NLP and then adapted for speech in HuBERT \cite{hsu2021hubert} and DiscreteBERT \cite{discretebert}, which consist of masking segments from the input signal and learning to predict discrete targets corresponding to the masked segments.
In HuBERT the discrete targets are obtained from k-means clustering of MFCCs and, in a second stage, from internal representations. \leo{BEATs \cite{chen2023beats} uses a similar approach as HuBERT, but the outputs of a random quantizer are used as targets and iteratively refined.}
In this work we propose to use the discrete units learned by EnCodec \cite{defossez2022high}, a general audio neural codec, as targets for our model, 
and then introduce, as in HuBERT, an additional target by clustering internal representations.
We use the Masked Autoencoder (MAE) architecture \cite{he2022masked}, which enables efficient pretraining.
Our proposed model, EnCodecMAE is novel in several aspects: (i) \leo{a neural audio codec (EnCodec) is used to generate discrete targets to learn universal audio representations;} 
(ii) a MAE architecture is used with frames instead of rectangular patches, achieving a performance in environmental tasks comparable to the best patch-based MAEs in the literature, while providing higher time resolution and improving performance in speech tasks.
\leo{(iii) as far as we know, this is the first work reporting the performance of a universal audio model in ASR.}
\leo{As shown in the results section, EnCodecMAE
achieves the best global performance when compared with various state-of-the-art  universal audio representations.}

\section{Proposed model}

\begin{figure}[t]
\centering
\centerline{\includegraphics[width=0.4\textwidth]{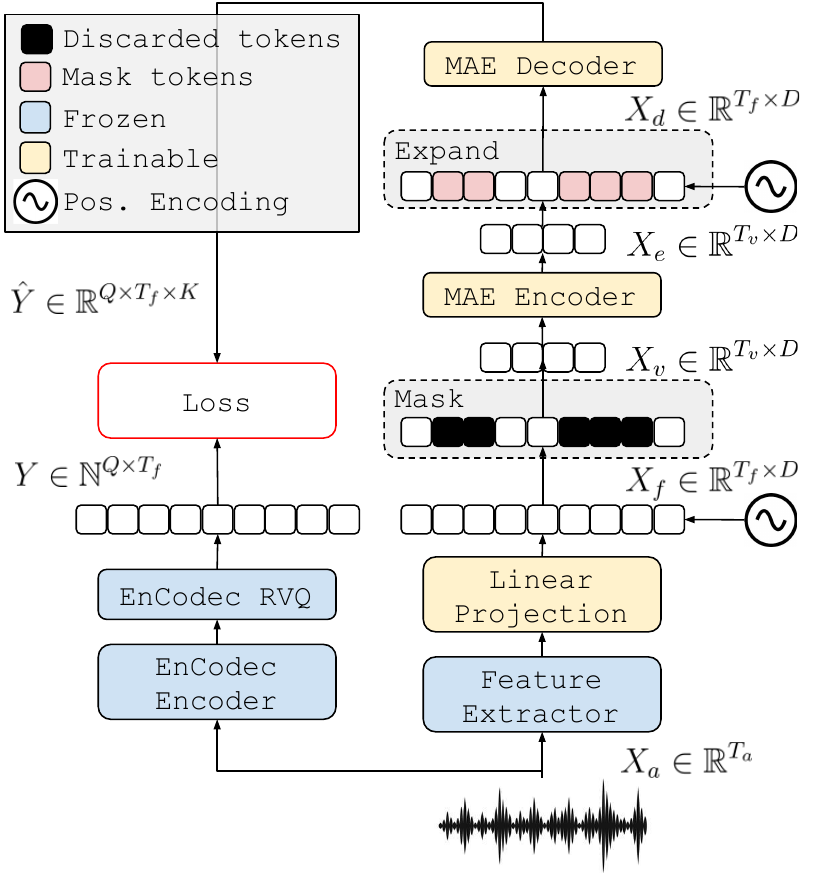}}
\vspace{-3mm}
\caption{EnCodecMAE architecture.
\leo{Features $X_f$ are extracted from an audio signal $X_a$, projected to the model dimensionality $D$, and positional embeddings are added.}
A percentage of the frames are masked and discarded, and the resulting sequence \leo{$X_v$} is processed by the MAE encoder. Before feeding the encoder output \leo{$X_e$} to the decoder, mask tokens are inserted in the positions that were dropped. The loss is finally computed between the posteriors \leo{$\hat{Y}$} generated by the MAE's decoder and the discrete targets \leo{$Y$} produced by \leo{EnCodec}.}

\label{fig:EnCodecmae}
\end{figure}

In this work we propose EnCodecMAE, a masked autoencoder that 
\leo{uses EnCodec representations as targets during learning.}
The EnCodecMAE architecture is depicted in Figure \ref{fig:EnCodecmae}. 
\leo{Features $X_f$ are extracted from an audio  signal $X_a$ with length $T_a$. We explored the use of melspectrograms and outputs from the EnCodec encoder as input features. In both cases, 75 embeddings per second are generated, resulting in a sequence of length~$T_f$.}
Similarly to wav2vec 2.0 and HuBERT, a proportion $M_\text{prop}$ of the embeddings is randomly masked, sampling the starting indices for the mask without replacement, and masking the subsequent $M_\text{gap}$ time steps for every sampled index. In contrast to wav2vec 2.0 and HuBERT, the masked embeddings are discarded from the sequence as in \cite{he2022masked}, instead of replaced by a mask token.
As a consequence, the resulting sequence \leo{$X_v$}, which contains only the visible embeddings, 
\leo{has a shorter length $T_v$ than the original one}. 
\leo{$X_v$} is passed through an encoder and then expanded to the original size \leo{$T_f$} by inserting mask tokens in the masked regions. Sinusoidal positional embeddings are added again to inform the decoder of the position of the masked regions that need to be reconstructed, and the resulting sequence \leo{$X_d$} is input to a decoder.
The model is trained to minimize a weighted cross entropy loss between the posteriors \leo{$\hat{Y}$} output by the decoder and the discrete target, \leo{$Y$}, generated by the EnCodec residual vector quantization (RVQ) layer. The following sections explain the components of the model. Source code is available at \url{https://github.com/habla-liaa/encodecmae}.

\subsection{EnCodec}

EnCodec \cite{defossez2022high} is a neural audio codec consisting of an 
\leo{autoencoder} with a quantized latent space. EnCodec achieves a high compression rate while also minimizing the perceptible distortion of the decoded audios, obtaining higher MUSHRA scores than hand-crafted codecs and Soundstream \cite{zeghidour2021soundstream} using the same bitrate,
both for speech and music signals.
This suggests that the quantized latent space from EnCodec contains most of the perceptually-relevant information for reconstructing not only speech, but also more complex signals like music.

The EnCodec architecture consists of a CNN encoder with 4 convolutional blocks that downsample the audio signal from 24kHz to 75 Hz, followed by 2 LSTM layers and a final convolutional layer. The encoder's output is a sequence of 128-dimensional vectors, which are passed through a residual vector quantization (RVQ) block with 32 codebooks, each with 1024 possible values. RVQ maps the input to the closest entry in the first codebook. Then, the residual is computed and mapped to the closest entry of the second codebook. This process is repeated for every codebook, increasingly refining the quantized outputs. 
Finally, the decoder mirrors the encoder architecture, upsampling the encoded signal with transposed convolutions and recovering the original audio.

\subsection{Masked autoencoders}
Masked autoencoders (MAE) were proposed in \cite{he2022masked} as a self-supervised model for computer vision and an alternative to Vision Transformers (ViT), and have been recently applied to audio representation learning \cite{huang2022amae, baade2022mae, chong2023masked}, using spectrogram patches as input and as targets during training.
They are trained in a similar way to BERT, by masking parts of the input signal and predicting the masked parts from the unmasked ones.
In contrast to BERT, which replaces the masked regions with a mask token, MAEs discard the masked regions  before feeding the sequence to the encoder. This makes training more efficient by reducing the length of the sequence.
As usually done with transformers, positional embeddings are added to the encoder input. Importantly, this is done before masking so that the remaining embeddings carry information on their absolute location in the sequence.
The encoder and decoder networks consist of a stack of transformer layers.
Before decoding, the encoder's output is expanded to its original length by adding mask tokens learned during training as a shared embedding.
The decoder
has a final classification layer with softmax activation. This layer predicts the discrete targets from the EnCodec's RVQ, corresponding to neural codec representations of the input audio. This is subtly different from the original MAE as the decoder is not designed to output an estimate of the encoder's input before masking, but a discretized version of it.
We use the assymetric encoder-decoder architecture proposed in \cite{he2022masked}, with a small decoder and a larger encoder.
In our setup, for the encoder we use 5 transformer layers in the small model, 10 in the base, and 20 in the large. All our models use only 2 layers in the decoder as in \cite{he2022masked} \leo{and pre-post layer normalization \cite{xie2023residual}}.
The dimension of the embeddings \leo{$D$} is 768 for the small and base models, and 1024 for the large model.

\subsection{Training procedure}

The training procedure consists of two stages.
Initially, the targets for the model are generated by the EnCodec RVQ layer. These consist of an index $y_q(t)$ between 1 and 1024 for each time step~$t$ and codebook $q$. 
The loss function is a weighted cross-entropy, with tunable weights for masked versus unmasked time steps and each individual codebook:
\begin{equation}
    L = \sum_{q=0}^{Q} \gamma_q \left[ \alpha \sum_{t \in M} C(y_q(t), \hat{y}_q(t)) + \beta \sum_{t \notin M} C(y_q(t), \hat{y}_q(t))\right]
\label{eq:loss} \nonumber
\end{equation}
where $\hat{y}_q(t)$ is the vector of 1024 posteriors generated by the decoder for time step $t$ and codebook $q$, $M$ is the set of masked indices, and $\alpha = \delta/|M|$ and $\beta = (1-\delta)/(T_f-|M|)$ are weights given to the masked and unmasked regions, respectively, with $\delta$ being a tunable parameter. 
The loss corresponding to each codebook stream $q$ is, in turn, weighted by $\gamma_q$. The loss for each codebook and time step is given by $C(y_q(t), \hat{y}_q(t)) = -\log(\hat{y}_{q,y_q(t)}(t))$, i.e., the negative logarithm of $\hat{y}_q(t)$ evaluated at the corresponding target index. 
Finally, after 500k training steps, a self-training stage is performed. We extract embeddings from the last encoder layer for 10k randomly-sampled audio signals and train a k-means model with $k=1024$. 
\leo{The model is trained for 150k additional steps to predict the discrete targets generated by the k-means model.}
As in HuBERT, for the large model we perform k-means on activations from the base model after self-training.

\section{Experimental setup}
\subsection{Pretraining}
For pretraining we used a mixture of datasets composed of Audioset (AS) \cite{audioset}, Free Music Archive \leo{(FMA)} \cite{fma_dataset} and the 6000 hours Libri-Light split (LL6K) \cite{librilight}. Audioset contains over 2 million 10-second audio clips from YouTube videos covering 527 different audio events. 
We downloaded more than 1.6M audios, which amounts to around 4500 hours.
Free Music Archive consists of 106,574 30-second tracks covering 161 genres for a total of over 800 hours of music. 
Libri-Light consists of english speech from audiobooks.

In all our experiments we trained with 4-second segments by randomly cropping the audio samples. At inference time, longer audios are chunked to 4 seconds and the resulting embeddings are concatenated. 
\leo{Melspectrograms are calculated using a Hanning window with 640 samples and a hop size of 320 samples. These hyperparameters are chosen to match EnCodec rate. We experimented using 128 and 256 mel bins.}

Our models were trained for 500k steps using AdamW optimizer with a weight decay of 0.05, $\beta_1$ of $0.9$, $\beta_2$ of  $0.95$, a fixed learning rate of 1e-4, and a batch size of 128. EnCodec's parameters were downloaded from Hugging Face\footnote{\scriptsize{\url{https://huggingface.co/facebook/encodec_24khz}}} and remained unchanged during training.
We used 8 of the 32 quantizers available in EnCodec as targets for training. 
We selected the training hyperparameters based on a base model's performance on HEAREval's validation data, after 150k training steps. This process resulted in the following values: $\delta=0.9$, $M_\text{prop}=0.5$ and $M_\text{gap}=15$. 
We set each weight $\gamma_q$ to be the average quantization error for the $q$-th codebook computed over 150 random training samples, normalized to sum to 1.0 over all codebooks. This gave a small improvement on validation data compared to using uniform $\gamma_q$ values. For feature extraction, we found that the last layer of the encoder gave the best results on validation data.

\subsection{Downstream evaluation}
\label{sec:downstream}
We evaluated our models following the HEAREval procedure~\cite{turian2022hear} and a subset of its instance-level tasks. The embedding for each instance is obtained by averaging the frame-level activations from the last encoder layer.
The resulting embedding is fed to a multilayer perceptron and a grid search over hyperparameters is performed as described in \cite{turian2022hear}. The upstream model is frozen, so only the downstream model parameters are trained for each evaluation task.
For music, we evaluate in the 50-hour subset of the NSynth Pitch \cite{engel2017neural} dataset (NS), which consists of instrument notes. The goal of this task is to classify each sound into 88 possible pitch values. We also evaluate on music genre classification (GC), using the GTZAN Genre dataset \cite{tzanetakis2002musical}, which consists of 1000 30-second audio tracks categorized into 10 genres. For speech, we evaluate on the Google Speech Commands dataset v2 (SC) \cite{warden2018speech}, where each utterance has to be classified into 10 possible commands or noise or unknown, and on the CREMA-D dataset \cite{cao2014crema} for emotion recognition (ER) into 6 categories.
For environmental sounds, we evaluate on the Freesound Dataset 50K (FSD) \cite{fonseca2021fsd50k}, which contains 100 hours of labeled sound events belonging to one or more  of 200 possible classes from the Audioset ontology. We also evaluate in the ESC-50 dataset (ESC) \cite{piczak2015dataset}, which is a collection of 2000 audio samples each with a single acoustic event belonging to one of 50 possible classes.

We report accuracy for all tasks except FSD, where mean average precision (mAP) is used.
\leo{We also report a global metric as in \cite{turian2022hear} by normalizing the performance on each task, clipping the resulting value to [-1,1] and then taking the average of these values across tasks. The normalization is done by subtracting the mean and dividing by the standard deviation of the performance of each task on the HEAREval 2021 benchmark results. A value of +1 means that the performance of the system is at least one standard deviation better than the average performance on every evaluated task.}

\leo{We also evaluated our models in ASR following the SUPERB protocol \cite{yang2021superb}, which consists of a vanilla 2-layer 1024-unit BiLSTM optimized by CTC loss on characters. This downstream model receives sequences of embeddings from each EnCodecMAE layer and learns a weighted average to combine them into a unique input sequence. We halved the frame-rate of our embeddings by averaging every 2 contiguous embeddings to speedup the training. The model is trained in Librispeech 100-hours of clean speech and decoding is done using the Librispeech official 4-gram language model.}

\section{Results}
\label{sec:results}

\begin{table*}
\footnotesize
\centering
\begin{tabular}{l|cc|cc|cc|c|cl}
\toprule
 & \multicolumn{2}{c|}{Music} & \multicolumn{2}{c|}{Speech} & \multicolumn{2}{c|}{Env} & Global & Params(M) & Pretrain Data \\
 & NS & GC & SC & ER & FSD & ESC &  &  &  \\
\midrule
(1) Mel256→EC (Small)&\meanerror{85.9}{1.2}&\meanerror{\bfseries 87.5}{1.9}&\meanerror{95.2}{0.7}&\meanerror{74.8}{0.9}&\meanerror{48.5}{1.2}&\meanerror{80.1}{1.7}&94.8&43.6& Mixture \\
(2) EC→EC (Base)&\meanerror{\bfseries 91.7}{1.0}&\meanerror{85.6}{1.9}&\meanerror{92.2}{1.0}&\meanerror{72.0}{1.0}&\meanerror{44.1}{1.3}&\meanerror{74.6}{2.1}&83.4&94.0& Mixture \\
(3) Mel128→EC (Base)&\meanerror{84.6}{1.2}&\meanerror{86.8}{1.7}&\meanerror{96.1}{0.7}&\meanerror{\bfseries 76.5}{0.9}&\meanerror{49.4}{1.1}&\meanerror{79.0}{2.3}&95.3&86.6& Mixture \\
(4) Mel256→EC (Base)&\meanerror{84.6}{1.0}&\meanerror{\bfseries 87.6}{1.9}&\meanerror{96.3}{0.5}&\meanerror{75.5}{1.0}&\meanerror{49.5}{1.1}&\meanerror{79.8}{1.8}&95.9&86.6& Mixture \\
(5) \phantom{Mel}AS only&\meanerror{86.1}{1.1}&\meanerror{87.1}{1.9}&\meanerror{93.6}{0.7}&\meanerror{73.8}{1.1}&\meanerror{51.7}{1.3}&\meanerror{81.5}{2.1}&97.0&86.6& AS \\
(6) \phantom{Mel}FMA only&\meanerror{87.7}{1.1}&\meanerror{\bfseries 89.2}{2.0}&\meanerror{87.4}{0.9}&\meanerror{68.7}{1.1}&\meanerror{48.5}{1.3}&\meanerror{80.4}{2.2}&89.1&86.6& FMA \\
(7) \phantom{Mel}LL6K only&\meanerror{83.4}{1.3}&\meanerror{83.4}{2.0}&\meanerror{95.4}{0.6}&\meanerror{75.6}{1.1}&\meanerror{41.3}{1.2}&\meanerror{68.8}{1.6}&75.3&86.6& LL6K \\
(8) Mel256→EC (Base + ST)&\meanerror{81.7}{1.4}&\meanerror{\bfseries 87.8}{2.1}&\meanerror{\bfseries 96.9}{0.5}&\meanerror{\bfseries 76.3}{1.0}&\meanerror{50.7}{1.2}&\meanerror{81.6}{2.1}&97.6&86.6& Mixture \\
(9) \phantom{Mel}AS only&\meanerror{83.5}{1.3}&\meanerror{\bfseries 87.7}{2.3}&\meanerror{94.6}{0.6}&\meanerror{75.1}{1.1}&\meanerror{\bfseries 52.9}{1.1}&\meanerror{\bfseries 84.6}{1.6}&99.2&86.6& AS \\
(10) Mel256→EC (Large)&\meanerror{85.3}{1.4}&\meanerror{85.8}{2.2}&\meanerror{96.4}{0.5}&\meanerror{\bfseries 77.1}{0.9}&\meanerror{50.3}{1.1}&\meanerror{80.2}{1.9}&97.1&261.6& Mixture \\
(11) Mel256→EC (Large + ST)&\meanerror{83.4}{1.0}&\meanerror{87.2}{2.1}&\meanerror{\bfseries 97.0}{0.4}&\meanerror{\bfseries 77.2}{0.9}&\meanerror{51.0}{1.0}&\meanerror{84.1}{1.8}&\bfseries 100.0&261.6& Mixture \\
(12) \phantom{Mel}AS only&\meanerror{86.4}{1.3}&\meanerror{86.4}{1.7}&\meanerror{94.4}{0.7}&\meanerror{75.7}{0.9}&\meanerror{\bfseries 53.4}{1.0}&\meanerror{\bfseries 86.0}{1.8}&99.1&261.6& AS \\
\midrule
(13) BEATS (Iter 1) \cite{chen2023beats} &\meanerror{82.4}{1.3}&\meanerror{85.4}{2.4}&\meanerror{91.0}{0.8}&\meanerror{68.7}{1.1}&\meanerror{50.6}{0.9}&\meanerror{82.3}{1.7}& 94.4 & 90 & AS \\
(14) BEATS (Iter 3) \cite{chen2023beats} &\meanerror{82.9}{1.0}&\meanerror{86.2}{2.9}&\meanerror{90.4}{0.8}&\meanerror{69.0}{1.1}&\meanerror{\textbf{53.5}}{1.1}&\textbf{\meanerror{85.1}{1.5}}& 96.0 & 90 & AS \\
(15) BYOL-A \cite{niizumi2021byol} &\meanerror{68.3}{1.4}&\meanerror{86.0}{2.0}&\meanerror{93.1}{0.7}&\meanerror{65.9}{1.1}&\meanerror{51.3}{0.9}&\meanerror{83.6}{2.0}& 82.6 & 5.3 & AS \\
(16) AudioMAE (PT) \cite{huang2022amae} &\meanerror{67.2}{1.5}&\meanerror{72.0}{3.1}&\meanerror{45.2}{1.6}&\meanerror{39.3}{1.0}&\meanerror{37.6}{1.2}&\meanerror{58.5}{2.1}& -32.1 & 86 & AS \\
(17) MSM-MAE-512 \cite{niizumi2022masked} $\dagger$ & 81.2 & 86.1 & 86.4 & 73.4 & 52.2 & \textbf{85.6} & 91.7 & 86 & AS \\
(18) Fuse HuBERT \cite{wu2022efficacy} $\dagger$ & 68.8 & 79.6 & 95.7 & 75.2 & 41.3 & 74.3 & 62.9 & 1000 & LL60K \\
(19) Fuse Wav2Vec2 \cite{wu2022efficacy} $\dagger$ & 60.6 & 79.3 & \textbf{96.9} & 69.2 & 40.3 & 69.5 & 51.4 & 317 & LV60K \\
\bottomrule
\end{tabular}
\vspace{3pt}
\caption{Performance in a subset of tasks from HEAREval benchmark. The first block corresponds to our models, the second to models from recent literature pretrained in an unsupervised way. We name our models with the syntax Input→Target, ie. Mel128→EC stands for a model that takes 128-bins melspectrograms as input and predicts EnCodec tokens. ST stands for self-training. We also report the global performance, number of parameters and datasets used to pretrain the models. LV60K and LL60K stand for 60k hours of LibriVox and Libri-Light respectively. The six datasets are described in section \ref{sec:downstream} 
Mixture consists of AS, FMA and LL6K.
For lines with a $\dagger$, we show the HEAREval leaderboard results or the ones reported in the original work. For BEATs and BYOL-A we used the official weights and adapted the implementation to be used with HEAREval.
We conducted bootstrapping with 100 iterations obtaining confidence intervals as the 2.5 and 97.5 percentiles \cite{ferrerconf}. Reported deviations are the difference between the metric and the furthest interval bound. The best performing model for each task, and those that fall within its confidence interval, are highlighted in bold text.}
\label{table:results-hear}
\end{table*}

Table \ref{table:results-hear} shows the results for various configurations of our proposed model and state-of-the-art models from recent literature, for the six tasks described in Section \ref{sec:downstream}. 

We compared different feature extractors and found that they have an important effect in the downstream performance and that the impact is different for each task. Globally, melspectrograms yielded better results than EnCodec representations (lines 2, 3 and 4 in Table \ref{table:results-hear}), and increasing the number of frequency bins from 128 to 256 slightly boosted the global performance (95.3\% vs 95.9\%, lines 3 and 4). However, for the task of pitch prediction (NS), our models using EnCodec outperformed melspectrograms and an ensemble of features (line 18) which concatenates embeddings from lines 16 and 17 and CREPE \cite{kim2018crepe}, a model trained for pitch prediction using annotated datasets.

\leo{We investigated the influence of the datasets used for pretraining on task performance (lines 4 through 7). Pretraining with speech-only data (LL6K Only) led to decreased performance in music and environmental tasks, while pretraining with music-only data (FMA Only) resulted in lower performance in speech tasks. Models pretrained on Audioset, or a mixture of Audioset, FMA, and Libri-Light, exhibited more balanced performance across music, speech, and environmental tasks. Pretraining with Audioset alone gave a higher performance in environmental tasks but performed slightly worse than the mixture in speech tasks.}

\leo{Scaling up the model size gave consistent improvements in global performance.}
\leo{Further, self-training (lines 8, 9, 11, and 12) enhanced our model's performance across most tasks. 
This finding aligns with prior research \cite{hsu2021hubert, chen2023beats}, reinforcing the effectiveness of self-training techniques in improving SSL representations.}

\leo{BEATs \cite{chen2023beats}, BYOL-A \cite{niizumi2021byol}, AudioMAE \cite{huang2022amae} and MSM-MAE-512 \cite{niizumi2022masked} are models pretrained in Audioset. 
Our base Audioset-trained model (line 5) outperforms these models on music and speech tasks. Our main differences with BEATs are the initial target definition (EnCodec in our model, random quantizers in BEATs), and the input signal format (frames in our model, rectangular patches in BEATs). 
Previous works \cite{niizumi2022masked,baade2022mae,gong2022ssast} found that using frames instead of rectangular patches gave better results in speech-related tasks but harmed event detection tasks. This is likely because rectangular patches span multiple frames and a frequency band, losing temporal resolution which might be useful for speech and tasks that require analysing a signal over time. Avoiding rectangular patching, which is borrowed from computer vision research \cite{dosovitskiy2020image}, and using frames instead, aligns better with traditional audio representations like spectrograms. Table 1 shows that our model reaches significantly better performance on speech tasks compared to patch-based models (compare line 5 with 13,16 and 17 and line 9 with 14) while performing on par with patch-based models in environmental tasks. This might be partially explained by our use of EnCodec instead of random quantizers or melspectrograms as targets. 
Performing self-training to enhance the initial targets, gives additional improvements (line 9).}

\begin{table}
\footnotesize
\centering
\begin{tabular}{l|ccc}
Model & w/o & w/LM & Params(M) \\
\toprule
Wav2Vec \cite{schneider2019wav2vec} & 15.86 & 11.00 & 32.5\\
VQ-Wav2Vec \cite{baevski2019vq} & 17.71 & 12.80 & 34.1\\
DeCOAR 2.0 \cite{ling2020decoar} & 13.02 & 9.07 & 89.8\\
Wav2Vec 2.0 Base \cite{baevski2020wav2vec} & 6.43 & 4.79 & 95.0 \\
Wav2Vec 2.0 Large \cite{baevski2020wav2vec} & 3.75 & 3.10 & 317.4 \\
HuBERT Base \cite{hsu2021hubert} & 6.42 & 4.79 & 94.7 \\
HuBERT Large \cite{hsu2021hubert} & 3.62 & 2.94 & 316.6\\
\midrule
Mel256→EC (Base) & 15.53 & 10.42 & 86.6 \\
\phantom{Mel} LL6K only & 14.91 & 10.12 & 86.6\\
Mel256→EC (Base+ST) & 14.41 & 9.90 & 86.6\\
\phantom{Mel} LL6K only & 13.72 & 9.29 & 86.6\\
Mel256→EC (Large+ST) & 12.44 & 8.59 & 261.6 \\
\end{tabular}
\vspace{3pt}
\caption{\leo{Word Error Rates (WER) obtained in the SUPERB ASR downstream task. We report metrics with and without a language model.}}
\vspace{-10pt}
\label{table:results-asr}
\end{table}

\leo{Fuse HuBERT and Fuse Wav2Vec 2.0 \cite{wu2022efficacy} generate embeddings by averaging the activations from all the transformer layers of HuBERT XL \cite{hsu2021hubert} and Wav2Vec 2.0 Large \cite{baevski2020wav2vec} models, respectively. These models are pretrained solely on speech data (60K hours of Libri-Light) with a focus in learning good representations for ASR.
Our base model trained with Libri-Light (line 7) performs better than these models across all non-speech tasks except for ESC where Fuse HuBERT is better. Our base model was able to perform similarly in speech-tasks in spite of having less parameters and being trained in a dataset an order of magnitude smaller. Moreover, training a large model with the mixture of datasets and self-training (line 11) gives better performance across all tasks, including SC and ER, while still being a smaller model trained on less data.}

\leo{Given the strong performance of our models in speech-related tasks, we further evaluated them in the more challenging task of ASR. The results in Table \ref{table:results-asr} indicate that our models still fall short of the performance achieved by the most recent speech SSL models such as HuBERT and Wav2Vec 2.0. Nevertheless, the performance is comparable to somewhat older SSL models like DeCOAR 2.0 and Wav2Vec.  
While pretraining only with speech data gives some improvements, it seems that the key to achieving HuBERT-level performance in ASR is not only in the dataset but also in the target definition, as clusters of MFCCs might be highly correlated with phonemes.}

\section{Conclusions}

We introduce EnCodecMAE, a novel universal audio representation model trained on masked inputs to predict targets from EnCodec, a neural audio codec. Our models, pretrained with Audioset or a mixture of datasets using self-training, achieve the best global performance across speech, music and environmental tasks. 
While pretraining only on Audioset results in a strong model, 
adding more speech leads to better results for speech-related tasks. 
Overall, our results show a clear trade-off between tasks depending on the input representation and the pretraining dataset. 
We also achieve a decent performance in ASR, comparable to DeCOAR 2.0. In future works we will focus in the challenging task of improving the performance in ASR without sacrificing performance in music and environmental tasks by exploring different training targets. 
EnCodecMAE is computationally efficient, with the largest model being trained in 5 days using 2 RTX 3090 GPUs. Further, given that our models are not fine-tuned and use light-weight downstream models, they allow for efficient implementation of multiple audio tasks in parallel with a single backbone.

\section{Acknowledgements}
This project has received funding from the European Union’s Horizon 2020 research and innovation programme under the Marie Skłodowska-Curie grant agreement No 101007666

\bibliographystyle{IEEEtran}
\bibliography{mybib}

\end{document}